\documentclass[twocolumn,groupedaddress,noshowpacs,nofootinbib]{revtex4}
\usepackage{graphicx}
\usepackage{amsmath}
\usepackage{epsfig}

\begin{document}

\title{SO(10) GUT Baryogenesis}

\author{Pei-Hong Gu$^{1}_{}$}
\email{pgu@ictp.it}

\author{Utpal Sarkar$^{2}_{}$}
\email{utpal@prl.res.in}

\affiliation{$^{1}_{}$The Abdus Salam International Centre for
Theoretical Physics, Strada Costiera 11, 34014 Trieste, Italy\\
$^{2}_{}$Physical Research Laboratory, Ahmedabad 380009, India}

\begin{abstract}

Baryogenesis, through the decays of heavy bosons, was considered to
be one of the major successes of the grand unified theories (GUTs).
It was then realized that the sphaleron processes erased any baryon
asymmetry from the GUT-baryogenesis at a later stage. In this paper,
we discuss the idea of resurrecting GUT-baryogenesis \cite{fy2002}
in a large class of SO(10) GUTs. Our analysis shows that fast lepton
number violating but baryon number conserving processes can
partially wash out the GUT-baryogenesis produced lepton and/or
baryon asymmetry associated with or without the sphaleron and/or
Yukawa interactions.

\end{abstract}

 \maketitle

One of the major triumphs of the GUTs is to \emph{naturally} explain
the observed matter-antimatter asymmetry \cite{pdg2006} in the
universe. In the GUT-baryogenesis \cite{yoshimura1978} scenario, the
quark-lepton unification provides the required baryon number
violation, the gauge and Yukawa couplings contain the required CP
phases and also satisfy the out-of-equilibrium condition near the
scale of grand unification. Thus the three conditions
\cite{sakharov1967} for baryogenesis are all satisfied and a baryon
asymmetry could be produced.

It was then pointed out \cite{thooft1976} that there is a
$\textrm{SU(2)}_L$ global anomaly in the standard model (SM), where
the baryon number $(B)$ and the lepton number $(L)$ are violated by
equal amount, so that their difference $B-L$ is still conserved.
Since anomalies are quantum processes that destroys classical
symmetries of any theory, this anomaly induced anomalous $B+L$
violating process will be suppressed by quantum tunneling
probability at zero temperature. But at finite temperature this
process becomes fast in the presence of an instanton-like solution,
the sphaleron \cite{krs1985}. During the period \cite{moore2000},
\begin{eqnarray}
100\,\textrm{GeV}\,\sim\,T_{EW}^{}\,< \,T\,
<\,T_{sph}^{}\,\sim\,10^{12}_{}\,\textrm{GeV}\,,
\end{eqnarray}
this sphaleron induced $B+L$ violating process will be too fast, and
hence, it will wash out any primordial $B+L$ asymmetry. However, it
will not affect any primordial $B-L$ asymmetry and will partially
transfer a $B-L$ asymmetry to a baryon asymmetry.

Note that $B-L$ is a global symmetry in the SU(5) GUTs, while it is
a local symmetry in the SO(10) or $\textrm{E}_6^{}$ GUTs. Thus the
baryon and lepton asymmetry generated in these theories at the GUT
scale (before the $B-L$ symmetry breaking) conserves $B-L$ and, in
fact, only a $B+L$ asymmetry is generated in all the GUTs. This
$B+L$ asymmetry would then be depleted exponentially by the
sphaleron transitions before the electroweak symmetry breaking, and
hence, the GUT-baryogenesis fails to explain why there are more
matter compared to antimatter in the universe.

To solve this problem of GUT-baryogenesis, many newer mechanisms
have been proposed. The most popular one is called leptogenesis
\cite{fy1986}, in which the lepton number violation required for the
neutrino masses generates a lepton asymmetry at some intermediate
symmetry breaking scale, which is then converted to a baryon
asymmetry in the presence of the sphalerons before the electroweak
phase transition. In this case the $B+L$ asymmetry generated by the
GUT-baryogenesis is washed out by the sphaleron transitions before
the leptogenesis begins. Recently, the resurrecting GUT-baryogenesis
\cite{fy2002} was proposed. In this scenario, the lepton number
violation starts before the sphaleron transitions begin, so that the
produced $B+L$ asymmetry in the GUT-baryogenesis is converted to a
$B-L$ asymmetry, which then generates the required baryon asymmetry
in the presence of the sphalerons.

In this paper, we discuss the idea of resurrecting GUT-baryogenesis
in the SO(10) GUTs. Our general analysis shows that in several
versions of SO(10) GUTs, it is possible to generate the
matter-antimatter asymmetry of the universe from GUT-baryogenesis.
Unlike the leptogenesis scenarios there is no necessity for the
lepton number violating interactions to generate a lepton asymmetry.
We start with the trivial case, when the SO(10) GUTs breaks down to
$SU(3)_c^{} \times SU(2)_L^{} \times U(1)_Y^{}$ at the GUT scale. In
this case $B-L$ is broken at the GUT scale and hence the asymmetry
generated by the decays of the heavy bosons is a $B-L$ asymmetry,
and hence the sphaleron transitions will not deplete it. However,
such single stage symmetry breaking SO(10) GUT scenario will have
all the problems of the SU(5) GUTs, like the gauge coupling
unification, fermion mass relations, strong CP problem and neutrino
masses. So, we shall exclude this possibility from the rest of our
discussions. We shall now present a popular version of the SO(10)
GUTs with conventional Higgs scalars and point out how the
GUT-baryogenesis in this model can explain the matter-antimatter
asymmetry of the universe without any added ingredients. This
mechanism can be applied to other versions of SO(10) GUTs, but we
shall not include them in our present discussion.

\vskip .2in

We start with an SO(10) GUT with the following symmetry breaking
pattern:
\begin{eqnarray}
SO(10) &\stackrel{M_{G}^{}}{\longrightarrow}&  SU(4)_{c}^{} \times
SU(2)_{L}^{} \times SU(2)_{R}^{}   \nonumber
\\
&\stackrel{M_{X}^{}}{\longrightarrow}&  SU(3)_{c}^{} \times
SU(2)_{L}^{} \times SU(2)_{R}^{} \times
U(1)_{B-L}^{}  \nonumber \\
&\stackrel{M_{B-L}^{}}{\longrightarrow}& SU(3)_{c}^{} \times
SU(2)_{L}^{} \times
U(1)_{Y}^{}   \nonumber \\
&\stackrel{m_{W}^{}}{\longrightarrow} & SU(3)_{c}^{} \times
U(1)_{em}^{} \,.
\end{eqnarray}
The unification of the gauge coupling constants constrains
the left-right symmetry breaking scale to be
greater than $10^{13}_{}\,\textrm{GeV}$.

In the present case we shall consider the following Higgs scalars for the
left-right and electroweak symmetry breaking:
\begin{eqnarray}
\hspace*{-1mm}\Delta_L^{} \hspace*{-0.7mm}\equiv\hspace*{-0.7mm}
(\textbf{1},\textbf{3},\textbf{1},-2)\hspace*{0.2mm},\hspace*{0.2mm}
\Delta_R^{} \hspace*{-0.7mm}\equiv\hspace*{-0.7mm}
(\textbf{1},\textbf{1},\textbf{3},-2)\hspace*{0.2mm},\hspace*{0.2mm}
\Phi \hspace*{-0.7mm}\equiv\hspace*{-0.7mm}
(\textbf{1},\textbf{2},\textbf{2},0)\hspace*{0.2mm}.
\end{eqnarray}
The left-right symmetry will be broken at the scale $M_{B-L}^{}$ by
the vacuum expectation value ($vev$) $\langle\Delta_R^{}\rangle$.
The left-right symmetry will ensure $M_{\Delta _L}^{}=M_{\Delta
_R}^{}\equiv M_{\Delta }^{}\sim \langle\Delta_{R}^{}\rangle$. The
bi-doublet Higgs can be looked on as a combination of two
$SU(2)_L^{}$ doublets in the SM, one denoted by $\phi$ corresponding
to the SM Higgs doublet having $vev$ $\langle\phi\rangle\simeq
174\,\textrm{GeV}$, while the other denoted by $\phi'$ having zero
$vev$. The $vev$ $\langle\Delta_L^{}\rangle$ is determined by the
minimization of the scalar potential and it is highly suppressed,
\begin{eqnarray}
\langle \Delta_L^{} \rangle = \alpha\, {\langle \phi \rangle^2 \over
\langle \Delta_R^{} \rangle}\,,
\end{eqnarray}
where $\alpha$ is some combination of
couplings entering in the scalar potential.

For simplicity, we only write down the following Yukawa couplings
and scalar interaction that are relevant for the rest of our discussions:
\begin{eqnarray}
\mathcal{L}& \supset &  -\,\frac{1}{2}\,f_{L_{ij}}^{}\,
\overline{\ell_{L_i}^{}}\,i\,\tau_{2}^{}\,\Delta_{L}^{}\,
\ell_{L_j}^{c}\,-\,\frac{1}{2}\,f_{R_{ij}}^{}\,
\overline{\ell_{R_i}^{}}\,i\,\tau_{2}^{}\,\Delta_{R}^{}\,
\ell_{R_j}^{c}
\nonumber\\
&&- \,y_{1_{ij}}^{} \,\overline{\ell_{L_i}^{}}
\,\Phi_{1}^{}\,\ell_{R_j}^{} \, -\,y_{2_{ij}}^{}\, \overline
{\ell_{L_i}^{}}\, \Phi_{2}^{}\,\ell_{R_j}^{}
\nonumber\\
&&-\,\gamma_{ij}^{}\,\textrm{Tr}\,\left(\,\Delta_{R}^{\dagger}\,\Phi_{j}^{}\,
\Delta_{L}^{}\,\Phi_{i}^{\dagger}\,\right)\,
+\,\textrm{h.c.}\,,
\end{eqnarray}
where $\ell_L^{} \equiv (\textbf{1},\textbf{2},\textbf{1},-1)$ and $
\ell_R^{} \equiv (\textbf{1},\textbf{1},\textbf{2},-1)$ are the
leptons, $\Phi_{1}^{}\equiv \Phi$ and $\Phi_{2}^{}\equiv
\tau_{2}^{}\Phi^{\ast}_{}\tau_{2}^{}$. After the left-right symmetry
is broken by $\langle\Delta_R^{}\rangle$, we deduce
\begin{eqnarray}
\label{leptonviolation} \mathcal{L} &\supset & -\,
\frac{1}{2}\,f_{L_{ij}}^{}\,
\overline{\ell_{L_i}^{}}\,i\,\tau_{2}^{}\,\Delta_{L}^{}\,
\ell_{L_j}^{c}\,-\,\mu\,
\phi^{T}_{}\,i\,\tau_{2}^{}\,\Delta_{L}^{}\,\phi
\nonumber\\
&& -\,y_{ij}^{}\,\overline{\ell_{L_i}^{}} \,\phi\,\nu_{R_j}^{}
\,-\,\frac{1}{2}\,M_{R_{i}}^{}\, \overline{\nu_{R_i}^{}}\,
\nu_{R_i}^{c} \,+\,\textrm{h.c.}\,,
\end{eqnarray}
with $\mu \sim \langle\Delta_{R}^{}\rangle$, $M_{R}^{}=
f_{R}^{}\langle\Delta_{R}^{}\rangle$. Here we have conveniently
chosen the basis, under which the Majorana mass matrix $M_{R}^{}$ is
diagonal and real. It is straightforward to see the second term
violates the \emph{left-handed lepton number} and the last term
violates the \emph{right-handed lepton number} by two units.

\vskip .2in

We now mention the general conditions that allows the
GUT-baryogenesis to explain the matter-antimatter asymmetry of the
universe:

\emph{Condition 1}: Without going into the details of symmetry
breaking at the GUT scale, we assume that a baryon asymmetry is
generated at this scale $M_G^{} \sim 10^{16}_{}\,\textrm{GeV}$
through the decays of some heavy bosons. Since $B-L$ is a local
symmetry below the GUT scale, the generated asymmetry is an $B+L$
asymmetry, which is expected in any GUT-baryogenesis.

\emph{Condition 2}: We demand that after the left-right symmetry is
broken, there will be fast lepton number violating interactions, but
there are no fast baryon number violating interactions. This is true
for most of the models. As discussed previously, the $vev$
$\langle\Delta_{R}^{}\rangle$ that breaks $U(1)_{B-L}^{}$ will lead
to the breaking of the lepton and baryon number. But since the
baryon number violation by two units requires higher dimensional
operators, these processes are never fast. For example, the simplest
dimension-9 operator is of the form $qqqqqq$, which corresponds to
three quarks going into three antiquarks. But this process can never
be in equilibrium since the probability of three quarks coming at a
point is strongly phase space suppressed and then this dimension-9
process is also suppressed by five powers of the heavy left-right
symmetry breaking mass scale. Thus, although
$\langle\Delta_R^{}\rangle$ breaks both baryon and lepton numbers,
only lepton number violation can be fast.

\emph{Condition 3}: Before the baryon and lepton asymmetry is
completely erased, the lepton number violating interactions will go
out of equilibrium. The sphaleron process directly affects the
left-handed fermions, but can also act on the right-handed fermions
if it is associated with the fast Yukawa interactions. In other
words we need to consider both left-handed and right-handed fermions
for the decoupling of the lepton number violating interactions, so
that the existing asymmetry is not washed out by the sphaleron
process.

We then give a general analysis in a class of SO(10) GUTs, in which
the GUT-baryogenesis continues to work and the produced $B+L$
asymmetry is not completely wiped out by the sphaleron process. The
different cases can be classified by the decoupling temperatures for
the left-handed lepton number $(L_L^{})$ and right-handed lepton
number $(L_R^{})$ violating interactions $T_L^{}$ and $T_R^{}$.

$\bullet$ For $T_L^{} > M_{B-L}^{}$ and $T_R^{} > T_{sph}^{}$, the
$L_R^{}$ asymmetry is washed out before the sphaleron transitions,
leaving the $B+L_L^{}$ asymmetry from the GUT-baryogenesis. This
$B+L_L^{}$ asymmetry gives a non-zero $B-L$, which then survives the
sphaleron process.

$\bullet$ Similarly, when $T_R^{} > M_{B-L}^{}$ and $T_L^{} >
T_{sph}^{}$, the $B+L_R^{}$ asymmetry remains unaffected and
contribute to a $B-L$ asymmetry.

$\bullet$ When the $L_L^{}$ and $L_R^{}$ violating interactions are
both in equilibrium before the sphaleron process becomes
operational, only the $B$ asymmetry from the GUT-baryogenesis
survives and contributes to a $B-L$ asymmetry, which then get
converted to a baryon asymmetry.

$\bullet$ If the $L_R^{}$ violating interactions remain out of
equilibrium, the $L_R^{}$ asymmetry may survive even after the
$B+L_L^{}$ asymmetry is erased by the joint sphaleron and Yukawa
interactions as long as the Yukawa interactions are not fast enough
to relate the number densities of the left- and right-handed leptons
at this stage. This $L_R^{}$ asymmetry can be related to the
$L_L^{}$ asymmetry after the Yukawa interactions become effective
and then can be converted to a baryon asymmetry.

\vskip .2in

For the purpose of demonstration, we consider a hierarchal case,
where $M_{\Delta,R}^{} >  M_{B-L}^{}\sim M_{W_R}^{}$. Here
$M_{W_{R}}^{}=g \langle\Delta_{R}^{}\rangle$ is the mass of the
charged right-handed gauge boson with $g$ being the $SU(2)$ gauge
coupling. Therefore, $\Delta_{L}^{}$ and $\nu_{R}^{}$ have decoupled
as the $B-L$ symmetry is broken. We thus integrate out the heavy
left-handed triplet Higgs and right-handed neutrinos from Eq.
(\ref{leptonviolation}) and then obtain a dimension-5 operator,
\begin{eqnarray}
\label{left} \mathcal{O}_{L}^{} &= & -\,\frac{1}{2}\,h_{ij}^{}\,
\overline{\ell_{L_i}^{}}\,\phi\,\phi^{T}_{}\,
\ell_{L_j}^{c}\,+\,\textrm{h.c.}\,
\end{eqnarray}
with $h=-y \displaystyle{\frac{1}{M_{R}^{}}}y^{T}_{}-
\displaystyle{\frac{\mu^{\ast}_{} }{M_{\Delta}^{2}}}f_{L}^{}$.
Obviously, this operator will give the neutrinos a small Majorana
mass matrix, $m_{\nu}^{}=h\langle \phi \rangle^{2}_{}$
\cite{minkowski1977} after the electroweak symmetry breaking.  Note
that the interaction (\ref{left}) only breaks the left-handed lepton
number. We can also obtain a $L_R^{}$ violating operator by
integrating out the right-handed neutrinos from the kinetic terms of
the right-handed leptons, for example,
\begin{eqnarray}
\label{right} \mathcal{O}_{R}^{} &= & -\,g_{ij}^{}\,
\overline{e_{R_i}^{}}\,W_{R}^{-}\,W_{R}^{-}\,
e_{R_j}^{c}\,+\,\textrm{h.c.}
\end{eqnarray}
with
\begin{eqnarray}
g_{ij}^{}\,=\,\frac{1}{2}\,g^{2}_{}\,V_{ik}^{}\frac{1}{M_{R_{k}}^{}}V_{kj}^{T}\,=\,\frac{1}{2}\,g^{2}_{}
\,\frac{1}{\tilde{M}_{R_{ij}}^{}}\,,
\end{eqnarray}
where $V$ is the orthogonal matrix diagonalizing the Majorana mass
matrix of the right-handed neutrinos. Since $ M_{W_R}^{} \sim
M_{B-L}^{}$, the induced $L_R^{}$ violating processes whose details
depend on the specific models will only remain effective for very
short interval of time and hence will not have a significant effect
on the GUT-baryogenesis produced baryon and lepton asymmetry.

For the left-handed lepton number violating processes,
$\ell_{L_i}^{}\phi^{\ast}_{}\leftrightarrow \ell_{L_j}^{c}\phi$ and
$\ell_{L_i}^{}\ell_{L_j}^{}\leftrightarrow \phi\phi$, the reaction
rate should be \cite{fy1990}
\begin{eqnarray}
\label{rateleft} \Gamma_{\not
L_L}^{}&=&\frac{1}{\pi^{3}_{}}\,T^{3}_{}\,\left|h_{ij}^{}\right|^{2}_{}\,
=\,\frac{1}{\pi^{3}_{}}\,\frac{T^{3}_{}\,\left|m_{\nu_{ij}}^{}\right|^{2}_{}}{\langle
\phi \rangle^{4}_{}}\,.
\end{eqnarray}
Comparing this reaction rate to the expansion rate of the universe,
\begin{eqnarray}
\label{hubble}
H&=&\left(\frac{8\,\pi^{3}_{}\,g_{\ast}^{}}{90}\right)^{\frac{1}{2}}_{}\,\frac{T^{2}_{}}{M_{\textrm{Pl}}^{}}\,
\end{eqnarray}
with $g_{\ast}^{}=\mathcal{O}(100)$ and $M_{\textrm{Pl}}^{}\simeq
1.22 \times 10^{19}_{}\,\textrm{GeV}$, we obtain,
\begin{eqnarray}
T_{L}^{}&\gtrsim
&\left(\frac{2^{2}_{}\,\pi^{9}_{}\,g_{\ast}^{}}{45}\right)^{\frac{1}{2}}_{}\,
\frac{\langle\phi\rangle^{4}_{}}{M_{\textrm{Pl}}^{}\,\left|m_{\nu_{ij}}^{}\right|^{2}_{}}\,\nonumber\\
\label{leftT} &\sim&
\left(\frac{\textrm{0.1\,eV}}{m_{\nu}^{}}\right)^{2}_{}\,\times\,3.9\,\times\,10^{12}_{}\,
\textrm{GeV}\,.
\end{eqnarray}

The right-handed leptons will be related to the left-handed ones
when their Yukawa interactions are in equilibrium. The rate of a
scattering process between the SM left- and right-handed fermions,
Higgs and $W$ bosons, $\psi_{L}^{}\phi\leftrightarrow
\psi_{R}^{}W^{-}_{}$, is
\begin{eqnarray}
\label{rateyukawa} \Gamma_{Y}^{}\sim \alpha_{W}^{}\lambda^{2}_{}T\,
\end{eqnarray}
with $\alpha_{W}^{}=g^{2}_{}/(4\pi)$ being the weak coupling
constant and $\lambda$ being the Yukawa couplings. Hence the Yukawa
interactions of electron, muon and tau leptons will keep in
equilibrium below the temperatures,
\begin{eqnarray}
T_{e}^{}\lesssim 10^{4}_{}\,\textrm{GeV}\,,\,T_{\mu}^{}\lesssim
10^{10}_{}\,\textrm{GeV}\,,\,T_{\tau}^{}\lesssim
10^{12}_{}\,\textrm{GeV}\,.
\end{eqnarray}

For illustration, let us consider a representative set of values for
the mass scales:
$M_{B-L}^{}\sim\mathcal{O}(10^{14}_{}\,\textrm{GeV})$ and
$m_{\nu}^{}=\mathcal{O}( 0.1\,\textrm{eV})$. The $L_L$ violating
interactions will decouple at $T_L \gtrsim 10^{12}\,\textrm{GeV}$.
In this case the GUT-baryogenesis generated a $B+L$ asymmetry at the
GUT scale $M_G^{}$. Below the $B-L$ breaking scale $M_{B-L}^{}$, the
$L_L^{}$ violating interactions are in equilibrium, but the
sphaleron and Yukawa processes have not yet started. So, although
the $L_L^{}$ asymmetry is depleted, the $B$ and $L_R^{}$ asymmetry
are not altered. But since $T_{L}^{} > T_{sph}^{}\sim T_{\tau}^{}$,
when the sphaleron process starts, the $L_L^{}$ violating
interactions will no longer be effective. Thus the $B+L_R^{}$
asymmetry generated by the GUT-baryogenesis will survive. In
consequence, a non-zero $B-L$ asymmetry will get converted to a
baryon asymmetry before the electroweak phase transition and then
give us the required matter-antimatter asymmetry.

\vskip .2in

We now discuss the possibility that the GUT-baryogenesis and
leptogenesis both contribute to the generation of the
matter-antimatter asymmetry of the universe. We consider two types
of $L_L^{}$ violating interactions: (a) the $L_L^{}$ violating
processes (\ref{left}) mediated by the left-handed triplet Higgs
$\Delta_L^{}$; (b) the $L_L^{}$ violating decays of
$N=\nu_R^{}+\nu_R^c$ to the left-handed leptons. For the proper
parameter choice, we have the flexibility to keep the (a) processes
in equilibrium before the sphaleron action becomes effective. Thus,
the GUT-baryogenesis produced $B+L_R^{}$ asymmetry will survive.
Subsequently, the $N$ can decay to generate a $L^{}_L$ asymmetry
above the temperature $T_{sph}^{}\sim T_{\tau}^{}$. When the
sphaleron interactions become effective, there will be no other
lepton number violating interactions, so the residual
$B+L_R^{}+L_L^{}$ asymmetry, where $B$ and $L_R^{}$ both come from
the GUT-baryogenesis while $L_L^{}$ comes from the leptogenesis,
will get converted to a $B-L$ asymmetry. This $B-L$ asymmetry can
give us a desired baryon asymmetry through the sphaleron process.

\vskip .2in

In this paper, we discuss the resurrecting GUT-baryogenesis in the
SO(10) GUTs. We did not attempt to present any specific model,
instead we restrict ourselves to making general statements comparing
the amplitudes of different processes with the expansion rate of the
universe. Since the parameter range for the different possibilities
are not too flexible, in a realistic model one should solve the
Boltzmann equation to get the exact numerical estimates of the
asymmetry. Our analysis shows that in a class of $\textrm{SO(10)}$
GUTs, after the $B+L$ asymmetry is produced by the decays of heavy
bosons, there are fast lepton number violating but baryon number
conserving interactions, which can partially wash out the existing
lepton and/or baryon asymmetry associated with or without the
sphaleron action and/or Yukawa interactions. Thus a $B-L$ asymmetry
can be generated from the GUT $B+L$ asymmetry. Subsequently, this
$B-L$ asymmetry gets converted to a baryon asymmetry in the presence
of the sphalerons before the electroweak symmetry breaking, which is
consistent with the observed matter-antimatter asymmetry of the
universe.

\end{document}